# Deep Active Learning for Lung Disease Severity Classification from Chest X-rays: Learning with Less Data in the Presence of Class Imbalance

Roy M. Gabriel[1], Mohammadreza Zandehshahvar[1], Marly van Assen[2], Nattakorn Kittisut[1], Kyle Peters[3], Carlo N. De Cecco[2], Ali Adibi[1]

[1]School of Electrical and Computer Engineering, Georgia Institute of Technology, Atlanta, GA, US
[2]Department of Radiology and Imaging Sciences, Emory University, Atlanta, GA, USA
[3]College of Computing, Georgia Institute of Technology, Atlanta, GA, USA

**Abstract**

To reduce the amount of required labeled data for lung disease severity classification from chest X-rays (CXRs) under class imbalance, this study applied deep active learning with a Bayesian Neural Network (BNN) approximation and weighted loss function. This retrospective study collected 2,319 CXRs from 963 patients (mean age, 59.2 ± 16.6 years; 481 female) at Emory Healthcare affiliated hospitals between January and November 2020. All patients had clinically confirmed COVID-19. Each CXR was independently labeled by 3 to 6 board-certified radiologists as normal, moderate, or severe. A deep neural network with Monte Carlo Dropout was trained using active learning to classify disease severity. Various acquisition functions were used to iteratively select the most informative samples from an unlabeled pool. Performance was evaluated using accuracy, area under the receiver operating characteristic curve (AU ROC), and area under the precision-recall curve (AU PRC). Training time and acquisition time were recorded. Statistical analysis included descriptive metrics and performance comparisons across acquisition strategies. Entropy Sampling achieved 93.7% accuracy (AU ROC, 0.91) in binary classification (normal vs. diseased) using 15.4% of the training data. In the multi-class setting, Mean STD sampling achieved 70.3% accuracy (AU ROC, 0.86) using 23.1% of the labeled data. These methods outperformed more complex and computationally expensive acquisition functions and significantly reduced labeling needs. Deep active learning with BNN approximation and weighted loss effectively reduces labeled data requirements while addressing class imbalance, maintaining or exceeding diagnostic performance.

**Statement and Declarations**

**Funding:**

This work was supported by the National Science Foundation (NSF) under the grant titled Collaborative Research: SCH: Intelligent Radiology Through Human-Machine Cooperation, ICOL# 149725.

**Competing Interests:**

The authors have no relevant financial or non-financial interests to disclose.

**Author Contributions:**

All authors contributed to the study conception and design. Material preparation, data collection and analysis were performed by all authors. The first draft of the manuscript was written by Roy Gabriel and all authors commented on previous versions of the manuscript. All authors read and approved the final manuscript.

# 1. Introduction

The recent pandemic and overall increasing prevalence and detection of pulmonary diseases have highlighted the urgent need for reliable and time efficient approaches for accurate diagnosis. Among these diagnostic solutions, chest X-rays (CXRs) remain a widely used imaging modality due to their availability, low cost, and diagnostic utility. However, the surge in imaging volume, in combination with a global shortage of radiologists, has led to delays in diagnosis and treatment [1].

The development of artificial intelligence (AI)-based tools has emerged as a critical solution, particularly in the medical field [2, 3]. However, AI models are highly dependent on the availability and quality of labeled training data, which is often limited by the significant expense and expertise required for labeling medical images [4]. Given these constraints, approaches like active learning are needed to reduce the number of required labels and to focus on gathering labels for the most informative images. Active learning is a machine learning technique that selectively chooses the most informative data points to train on. The key to effective active learning lies in the selection of samples, which is driven by an acquisition function. An acquisition function evaluates the informativeness of each unlabeled sample, guiding the selection process by identifying which samples, if labeled, would most improve the model. By doing this, active learning reduces the need for extensive annotated datasets while still achieving high performance [2, 3, 4, 5, 6, 7].

However, applying active learning in medical imaging presents its own set of challenges. Medical datasets are often imbalanced, where certain conditions, outcomes, or severity levels are underrepresented due to the complexity of medical images and the scarcity of data [8, 9]. This imbalance can complicate the learning process and lead to biased models that may not generalize well [10, 11].

This study investigates the integration of deep active learning techniques with a weighted loss strategy to handle class imbalance effectively. Using CXRs for the diagnosis of COVID-19 as a case study, we evaluate various acquisition functions to identify those most effective at achieving baseline diagnostic performance with minimal labeled data. We hypothesize that combining active learning with weighted loss training, which penalizes the loss function based on inverse class label proportions, will allow us to achieve competitive diagnostic performance with significantly less labeled training data.

Unlike prior work that assumes balanced datasets, overlooks imbalance, or proposes new acquisition strategies to mitigate the imbalance [12, 13, 14], we demonstrate that standard state-of-the-art (SOTA) acquisition functions, when paired with a weighted loss strategy, can effectively handle class imbalance without the need for specialized sampling mechanisms. The aim of this study is to determine whether this approach can achieve baseline diagnostic performance using only a small fraction of labeled training data. By doing so, we aim to offer a cost-effective and scalable approach for deploying AI in imbalanced medical imaging tasks.

## 2. Materials and Methods

### 2.1. Dataset

This retrospective study included 963 consecutive adult patients (≥18 years old) diagnosed with COVID-19 using real-time reverse transcriptase-polymerase chain reaction who underwent diagnostic CXRs at Emory Healthcare affiliated hospitals (Alanta, GA, USA) [15, 16, 17, 18]. The data included patients seen between 01/30/2020 and 11/05/2020. Of the 963 patients in this study, between 396 and 658 patients were included in prior publications that focused on different objectives, such as severity scoring or radiologist-AI variability assessment [15, 16, 17, 18]. This study expands upon that work by using a larger patient number and introducing an active learning framework to evaluate model performance across varying levels of class imbalance. The need for informed consent was waived by Emory University institutional review board. Both in-patients and out-patients were included. All CXRs acquired during the entire hospital admission of the patient were collected, ranging from 1 to 30 images per patient. Posteroanterior and anteroposterior CXR projections were obtained using standard clinical protocols. Lateral views were excluded due to limited standardization across patients. Depending on patient condition, CXRs were taken with a portable X-ray unit and in a supine, erect, or semi-erect position.

Basic demographic information, including age at the time of scan, gender, and body mass index (BMI), was collected. All patient data were de-identified according to Health Insurance Portability and Accountability Act and hospital-specific guidelines.

Data was labeled by three to six cardiothoracic radiologists, with experience reading CXRs ranging from two to fifteen years. Labels were divided into training, validation, and testing sets with a 70/10/20 split. The reference standard was determined by taking the median label from the radiologists for each CXR. None of the experts had insight into clinical data and/or outcomes and the data was presented in a randomized order. Disease severity was categorized as normal, moderate, or severe according to the guidelines outlined in Figure 1 (a-c). All readers used the same guidelines and were provided with examples for each class before labeling the dataset. The labeling was performed independently by the readers using an in-house designed labeling tool, which randomized the CXRs and allowed for image size adjustment for better visualization. No indeterminate results occurred.

*2.2. Data Preprocessing and Labeling*

Data preprocessing involved resizing, normalization, and augmentation of CXRs to prepare for effective model training. Images were resized to a standard dimension of $256^2$ pixels. There was no missing data. Augmentations included random horizontal flips and random affine transformations, such as translations, rotations, scaling, and shear transformations to enhance model robustness and generalization. For the binary classification task, we combined the moderate and severe classes, leaving a normal (14%) and an abnormal (86%) class. For the multi-class classification task, the class labels were normal (14%), moderate (36%), and severe (50%).

*2.3. AI Model and Training*

We utilized a ResNet50 architecture, incorporating Monte Carlo (MC) Dropout layers, which served as an approximation of Bayesian Neural Networks (BNNs) [19, 20, 21]. The final layers consisted of an MC Dropout layer, two linear transformations (512 units each), batch normalization, and ReLU activation, culminating in either softmax (multi-class) or sigmoid (binary) output layers. We used a learning rate of 0.001, the Adam optimizer, early stopping after 3 successive iterations, and ensured a stratified split across class labels to maintain the class label distribution across training, validation, and testing.

To mitigate the effects of class imbalance, we employed a weighted loss strategy during model training. The weighted loss function penalized the model based on the inverse of the class distribution. This weighted loss strategy ensured that the minority classes are given more weight in the learning process. This approach helped to counterbalance the skewed distribution of the dataset, allowing the model to better recognize and classify underrepresented classes.

*2.4. Active Learning Strategy*

The active learning framework iteratively expanded the training set, $D_{train}$ using acquisition functions to identify the most informative samples from an unlabeled pool, $D_{pool}$. The process was randomly initialized with 25 labeled CXRs per severity class from $D_{train}$, and in each subsequent iteration, 20 additional samples were acquired from $D_{pool}$ using an acquisition function and labeled. This iterative sampling continued until either the model reached baseline diagnostic performance or the $D_{pool}$ was exhausted. The active learning implementation logic can be seen in Figure 2. For more details, see the first section in Online Resource.

The baselines used in both the binary and multi-class settings are BNN models trained on the entire training dataset, which represents 70% of the total dataset. The results on the testing set were used as a benchmark.

*2.5. Acquisition Function*

Seven acquisition functions guided sample selection: Random Sampling, Entropy Sampling [22, 23], BatchBALD [24], Mean STD [25], Least Confidence [26], Margin Sampling [27], and Variation Ratios [28]. Detailed definitions and mathematical formulations are provided in the second section of the Online Resource.

*2.6. Evaluation Metrics*

We evaluated the performance of the model using accuracy, negative log-likelihood (NLL), F1 score, precision, sensitivity, specificity, area under the ROC curve (AU ROC), and area under the precision-recall curve (AU PRC). Computational efficiency was evaluated by measuring Training Model Elapsed Time and Batch Acquisition Elapsed Time. Experiments were executed using Python on an RTX 2080 GPU, repeating each acquisition function evaluation with ten distinct random seeds. We showcase the result using the median with interquartile ranges (IQR) using the 25th and 75th percentiles.

We also compared the sampling techniques with random sampling. The optimal acquisition function was identified based on the lowest median number of training data samples and the narrowest IQR across most evaluation metrics. For both classification tasks, we determined the percentage of training data at which this acquisition function first achieved baseline performance. This same subset of training data, selected by the optimal acquisition function, was subsequently used for random sampling as well. Evaluation metrics were computed using models trained on both the actively and randomly sampled subsets, and the mean and standard deviation were calculated.

The dashed line (-) in the result tables denotes that the active learning algorithm was terminated without achieving baseline performance. It is worth mentioning that if the random sampler did not reach the baseline metrics (represented by a dashed line), as it theoretically should have, it means that almost all of the training data was exhausted. If all of the data had been exhausted, the random sampler would have achieved baseline metric performance, since the same samples used in the baseline would have been used in the final iteration.

## 3. Results

The data included 2,199 CXRs from 963 patients, with the mean age of the patients being $59.2 \pm 16.6$ years and 49.9% ($n = 481$) of the patients being female. The average BMI was $32.7 \pm 10.9$. It is worth mentioning that 120 CXRs lacked demographic information and were omitted from Table 1. Of the labeled images, 1,157 were severe, 836 mild, and 326 normal.

### 3.1. Binary Results

The baseline results can be seen in Online Resource: Table 1. The results detailed in Figure 3 were achieved using a baseline accuracy of 93.75% as a stopping criterion for the active learning algorithm. Table 2 shows a summary of the binary classification results. For all metrics, see Online Resource: Figure 1.

Our findings indicate that Entropy Sampling performs the best in the binary setting, achieving baseline accuracy of around 94% with only 15.4% (IQR: [14.17, 27.73]) of the training data outperforming more complex acquisition functions like BatchBALD and Variation Ratios, which required 17.87% (IQR: [16.64, 28.96]) and 24.03% (IQR: [17.87, 27.73]) of the training data, respectively. Furthermore, our approach surpasses the baseline performance for AU ROC and AU PRC with less training data. In the binary setting, Entropy Sampling achieves an AU ROC of 94% and an AU PRC of 99%, surpassing the baseline AU ROC of 91% while matching the AU PRC baseline, using only 8.01% (IQR: [8.01, 11.71]) of the training data for AU ROC and only 11.71% (IQR: [8.01, 14.17]) of the training data for AU PRC. It is worth mentioning that Least Confidence achieved baseline AU PRC with less data albeit a wider IQR, 9.24% (IQR: [6.78, 16.64]). Of note, all acquisition functions outperform Random Sampling, confirming proper model implementation and expected performance.

### 3.2. Multi-Class Results

The baseline results can be seen in Online Resource: Table 2. After using a baseline accuracy of 70.25% as a stopping criterion for the active learning algorithm, we achieved the following results as seen in Figure 4. Further, Table 3 shows a summary of the multi-class classification results. For all metrics, see Online Resource: Figure 2.

Our findings indicate that Mean STD performs the best in a multi-class setting, achieving a baseline accuracy of 70% with only 23.11% (IQR: [19.41, 29.27]) of the training data, In the multi- class setting, Mean STD achieves an AU ROC of 87% and an AU PRC of 77%, slightly exceeding the baseline AU ROC of 86% and AU PRC of 75%, respectively, using only 23.11% (IQR: [19.41, 29.27]) and 19.41% (IQR: [19.41, 29.27]) of the training data, respectively. This sampling technique outperforms more complex acquisition functions like BatchBALD and Variation Ratios. BatchBALD required 26.80% (IQR: [23.11, 35.43]), 25.57% (IQR: [20.64, 28.03]), and 20.64% (IQR: [19.41, 25.57]) of the training data to achieve baseline performance for accuracy, AU ROC, and AU PRC, respectively. Variation Ratios required 25.57% (IQR: [23.11, 35.43]), 25.57% (IQR: [23.11, −]), and 28.03% (IQR: [24.34, −]) of the training data to achieve baseline performance for accuracy, AU ROC, and AU PRC, respectively. It is worth mentioning that Entropy Sampling achieved baseline AU ROC using less data albeit a slightly wider IQR, 19.41% (IQR: [16.94, 28.03]).

Like in the binary setting, all acquisition functions outperform Random Sampling, confirming proper model implementation and expected performance.

*3.3. Comparing Active Sampler with Random Sampling*

As seen in Figure 5 and Figure 6, relative to all metrics, Entropy Sampling (binary) and Mean STD (multi) outperform Random Sampling using the same subset of data, while still achieving and sometimes surpassing baseline level performance with minimal data used. The misclassification error of the random sampler is much higher than that of Entropy Sampling and Mean STD ($0.11 \pm 0.05$ vs. $0.06 \pm 0.01$ for Entropy Sampling and $0.33 \pm 0.03$ vs. $0.28 \pm 0.01$ for Mean STD), demonstrating the powerful results and consistency of our samplers. For all metrics, see Online Resource: Figure 3 and Figure 4.

*3.4. Sample Distribution and Computational Efficiency Across Acquisition Functions*

As shown in Online Resource: Table 3 and Figure 5, Random Sampling largely reflected the original class distribution of the dataset, 13.3% normal and 86.7% abnormal in the binary setting and 13.7%, 37.0%, and 49.3% for normal, moderate, and severe in the multi-class setting. In contrast,

Entropy Sampling and Mean STD exhibited a shift toward oversampling the minority class. Entropy Sampling selected 35.3% normal samples in the binary setting, while Mean STD sampled 30.5% in the multi-class setting, one of the highest among all samplers. In multi-class setting, Mean STD sampled 30.5% normal, 41.8% moderate, and 27.8% severe, shifting focus toward the moderate class and sampling fewer from the overrepresented severe class. All acquisition functions besides Random Sampling followed a similar trend, sampling more from the moderate and normal classes and less from severe.

Timing analysis (Online Resource Table 4 and Figure 6) showed that batch acquisition time was negligible for Random Sampling and highest for BatchBALD, averaging 60-70 seconds per iteration. Entropy Sampling and Mean STD maintained relatively low acquisition times (41-44 seconds) while still achieving strong performance. Training times remained broadly comparable across all acquisition functions (245-309 seconds), with Mean STD trending slightly faster on average.

*3.5. Optimizing for Each Metric*

A summary of the optimal acquisition function based on each desired metric is shown in Table 4. We can see that Entropy Sampling is optimal if, with minimum data, the use case is to maximize accuracy, F1, AU ROC, and sensitivity in the binary setting and F1, AU ROC, and precision in the multi-class setting. Mean STD, however, can be used to maximize precision in the Binary setting and accuracy, AU PRC, sensitivity, and specificity in the multi-class setting. Other sampling techniques, like Least Confidence, are also useful for maximizing specificity in the Binary setting.

**4. Discussion**

This study demonstrates that deep active learning can substantially reduce labeling requirements for COVID-19 CXR classification, particularly when using Entropy Sampling for binary and Mean STD for multi-class tasks. These acquisition functions achieved baseline diagnostic accuracy with only 15.4% and 23.1% of the training data. Compared to more complex methods like BatchBALD and Variation Ratios, these methods reduced labeling needs by up to 73% and 35%, respectively.

Entropy Sampling and Mean STD also exhibited narrower IQRs for accuracy, F1, and sensitivity, indicating consistent and reproducible performance. This is critical in clinical deployment where variability can undermine trust in AI. Additionally, their low acquisition latency further supports their practicality over methods like BatchBALD, which are computationally expensive and less consistent.

These findings show the practicality of combining simpler acquisition functions with a weighted loss strategy to manage class imbalance, a common issue in medical imaging datasets [8, 9, 29, 14, 30, 31, 32].

Previous work on label efficiency in radiology has often relied on more complex approaches. Wu et al. [13] achieved 86.6% accuracy with 42% labeled CTs using a loss prediction module and a balanced severity cohort. Mahapatra et al. [33] reached 95.3% AUROC using 35% of labeled CXRs with a combination of BNNs and generative adversarial networks. Nazir et al. [14] applied discriminative clustering to improve sampling under class imbalance. In contrast, our method achieved comparable or better performance with just 15.4-23.1% of labeled data, using standard acquisition functions and architectures, making it more scalable and easier to deploy in clinical settings.

Our approach is also flexible and adaptable in real-world settings, allowing different acquisition functions to be chosen based on clinical priorities. For example, Mean STD is preferable for precision focused binary tasks, whereas Entropy Sampling is preferred in multi-class settings when sensitivity is critical. Such adaptability is vital in scenarios like pandemics or rare disease contexts, where rapid deployment of diagnostic models is essential and labeling resources are scarce [34, 35].

There are several limitations in our study. The inter- and intra-reader variability is significant [16]. Achieving a more accurate ground truth may require sophisticated weighting methods for radiologist assessments, which requires further exploration. Another limitation is our study focuses exclusively on lung disease. Future research should examine our active learning approach on other highly imbalanced datasets across various medical domains and data modalities. Additionally, investigating patterns in selected images based on different acquisition functions may provide valuable insights into epistemic and aleatoric uncertainties, assisting radiologists with complex cases. Multi-center studies and varied patient data distributions should also be explored to evaluate the robustness and generalizability of our technique. Finally, our analysis of sampling patterns shows the operational viability of Entropy Sampling and Mean STD. Their tendency to oversample ambiguous or underrepresented classes mitigates class imbalance during learning. Overall, this simple yet effective approach reduces annotation burden and training cost without sacrificing performance, making it well-suited for scalable radiology AI workflows beyond COVID-19 and CXRs.

5. Conclusion

Deep active learning integrated with a BNN using MC dropout and weighted loss achieves competitive diagnostic accuracy on imbalanced CXR datasets using only 15.4% of the training data for binary classification and 23.1% for multi-class classification. This significantly reduced the amount of data needed to train AI diagnostic models and radiologists' data labeling requirements. Among seven sampling techniques from naive techniques like Random Sampling to advanced and complicated techniques like BatchBALD, Entropy Sampling and Mean STD emerged as the most effective techniques, outperforming more complex and computationally expensive acquisition methods by selectively targeting the most informative samples. This study offers the opportunity to decrease the cost and amount of data annotation requirement, accelerating the deployment of AI-driven diagnostic tools. It supports radiologists in labeling the least amount of data needed to achieve the highest performance metric they desire, such as AU ROC, sensitivity, and/or specificity.

**Tables**

Table 1: Statistics and patient characteristics for the CXR datasets used in the study.

| Characteristic     | Statistic      |
|--------------------|----------------|
| Number of Images (N) | 2,199        |
| Patients           | 963            |
| Age (mean ± SD)    | 59.2 ± 16.6    |
| BMI (mean ± SD)    | 32.7 ± 10.9    |
| Female Count       | 481 (49.9%)    |

Table 2: Binary classification results showing percentage of training data used until baseline performance was reached. The optimal median number of imaging data samples are represented in bold for each metric, and the corresponding acquisition function would be deemed optimal. The interquartile range (IQR) column represents the 25th and 75th percentiles. The dashed line (-) denotes that the active learning algorithm was terminated without achieving baseline performance.

| Acquisition Function | Metric | Median % of Training Data | IQR |
|---|---|---|---|
| Entropy Sampling | Accuracy | **15.40** | [14.17, 27.73] |
| | F1 | **14.17** | [12.94, 16.64] |
| | AU PRC | 11.71 | [8.01, 14.17] |
| | AU ROC | **8.01** | [8.01, 11.71] |
| | Precision | - | [17.87, -] |
| | Sensitivity | **5.55** | [4.31, 10.47] |
| | Specificity | 22.80 | [15.40, -] |
| BatchBALD | Accuracy | 17.87 | [16.64, 28.96] |
| | F1 | 16.64 | [12.94, 24.03] |
| | AU PRC | 10.47 | [9.24, 17.87] |
| | AU ROC | 10.47 | [8.01, 15.40] |
| | Precision | 24.03 | [14.17, -] |
| | Sensitivity | 8.01 | [6.78, 9.24] |
| | Specificity | 28.96 | [15.40, -] |
| Random Sampling | Accuracy | 57.30 | [32.66, -] |
| | F1 | 37.58 | [17.87, 46.21] |
| | AU PRC | 17.87 | [15.40, 36.35] |
| | AU ROC | 12.94 | [5.55, 20.33] |
| | Precision | - | [26.49, -] |
| | Sensitivity | 15.40 | [5.55, 19.10] |
| | Specificity | 42.51 | [36.35, -] |
| Least Confidence | Accuracy | 19.10 | [17.87, 25.26] |
| | F1 | 16.64 | [10.47, 20.33] |
| | AU PRC | **9.24** | [6.78, 16.64] |
| | AU ROC | **8.01** | [6.78, 10.47] |
| | Precision | - | [16.64, -] |
| | Sensitivity | 6.78 | [4.31, 9.24] |
| | Specificity | **15.40** | [12.94, -] |
| Variation Ratios | Accuracy | 24.03 | [17.87, 27.73] |
| | F1 | 17.87 | [10.47, 22.80] |
| | AU PRC | 11.71 | [10.47, 16.64] |
| | AU ROC | 10.47 | [5.55, 11.71] |
| | Precision | 20.33 | [16.64, -] |
| | Sensitivity | 6.78 | [5.55, 10.47] |
| | Specificity | 22.80 | [16.64, -] |
| Mean STD | Accuracy | 17.87 | [15.40, 24.03] |
| | F1 | 15.40 | [5.55, 19.10] |
| | AU PRC | 11.71 | [11.71, 17.87] |
| | AU ROC | 10.47 | [9.24, 12.94] |
| | Precision | **14.17** | [12.94, -] |
| | Sensitivity | 6.78 | [5.55, 15.40] |
| | Specificity | 22.80 | [15.40, -] |
| Margin Sampling | Accuracy | 19.10 | [12.94, 26.49] |
| | F1 | 16.64 | [12.94, 19.10] |
| | AU PRC | **9.24** | [8.01, 12.94] |
| | AU ROC | 9.24 | [5.55, 15.40] |
| | Precision | 15.40 | [11.71, -] |
| | Sensitivity | 9.24 | [6.78, 11.71] |
| | Specificity | 17.87 | [10.47, -] |

Table 3: Multi-class classification results showing percentage of training data used until baseline performance was reached. The optimal median number of imaging data samples are represented in bold for each metric, and the corresponding acquisition function would be deemed optimal. The interquartile range (IQR) column represents the 25th and 75th percentiles. The dashed line (-) denotes that the active learning algorithm was terminated without achieving baseline performance.

| Acquisition Function | Metric | Median % of Training Data | IQR |
|---|---|---|---|
| Entropy Sampling | Accuracy | 26.80 | [18.18, 40.36] |
| | F1 | **20.64** | [18.18, 46.52] |
| | AU PRC | **19.41** | [16.94, 28.03] |
| | AU ROC | **19.41** | [16.94, 28.03] |
| | Precision | **19.41** | [15.71, -] |
| | Sensitivity | 18.18 | [14.48, 19.41] |
| | Specificity | 21.87 | [18.18, 34.20] |
| BatchBALD | Accuracy | 26.80 | [23.11, 35.43] |
| | F1 | 26.80 | [19.41, 30.50] |
| | AU PRC | 20.64 | [19.41, 25.57] |
| | AU ROC | 25.57 | [20.64, 28.03] |
| | Precision | - | [29.27, -] |
| | Sensitivity | 23.11 | [16.94, 26.80] |
| | Specificity | 23.11 | [19.41, 26.80] |
| Random Sampling | Accuracy | 35.43 | [29.27, 48.98] |
| | F1 | 46.52 | [31.73, -] |
| | AU PRC | 21.87 | [20.64, 30.50] |
| | AU ROC | 30.50 | [26.80, 31.73] |
| | Precision | - | [29.27, -] |
| | Sensitivity | 21.87 | [15.71, 30.50] |
| | Specificity | 28.03 | [21.87, 31.73] |
| Least Confidence | Accuracy | 28.03 | [21.87, 37.89] |
| | F1 | 31.73 | [21.87, 37.89] |
| | AU PRC | **19.41** | [16.94, 28.03] |
| | AU ROC | 21.87 | [19.41, 29.27] |
| | Precision | - | [40.36, -] |
| | Sensitivity | 16.94 | [16.94, 21.87] |
| | Specificity | 23.11 | [18.18, 31.73] |
| Variation Ratios | Accuracy | 25.57 | [23.11, 35.43] |
| | F1 | 25.57 | [23.11, 30.50] |
| | AU PRC | 28.03 | [24.34, -] |
| | AU ROC | 25.57 | [23.11, -] |
| | Precision | 30.50 | [23.11, -] |
| | Sensitivity | 18.18 | [18.18, 30.50] |
| | Specificity | 24.34 | [19.41, 30.50] |
| Mean STD | Accuracy | **23.11** | [19.41, 29.27] |
| | F1 | 23.11 | [19.41, 29.27] |
| | AU PRC | **19.41** | [19.41, 29.27] |
| | AU ROC | 23.11 | [19.41, 29.27] |
| | Precision | 24.34 | [19.41, -] |
| | Sensitivity | **15.71** | [14.48, 19.41] |
| | Specificity | **19.41** | [15.71, 29.27] |
| Margin Sampling | Accuracy | 24.34 | [19.41, 34.20] |
| | F1 | 24.34 | [23.11, 34.20] |
| | AU PRC | 24.34 | [19.41, 29.27] |
| | AU ROC | 26.80 | [19.41, -] |
| | Precision | 31.73 | [23.11, -] |
| | Sensitivity | **15.71** | [14.48, 28.03] |
| | Specificity | 23.11 | [16.94, 42.82] |

Table 4: Optimal acquisition function based on desired metric.

| Metric | Binary | Multi |
|---|---|---|
| Accuracy | Entropy Sampling | Mean STD |
| F1 | Entropy Sampling | Entropy Sampling |
| AU PRC | Margin Sampling* | Mean STD*** |
| AU ROC | Entropy Sampling** | Entropy Sampling |
| Precision | Mean STD | Entropy Sampling |
| Sensitivity | Entropy Sampling | Mean STD**** |
| Specificity | Least Confidence | Mean STD |

*Notes:*

\* *Least Confidence has the same median number of samples but a larger variance across trials.*

\*\* *Least Confidence has the same median number and a similar variance across trials.*

\*\*\* *Entropy Sampling and Least Confidence have the same median number of samples but a larger variance across trials.*

\*\*\*\* *Margin Sampling has the same median number of samples but a larger variance across trials.*

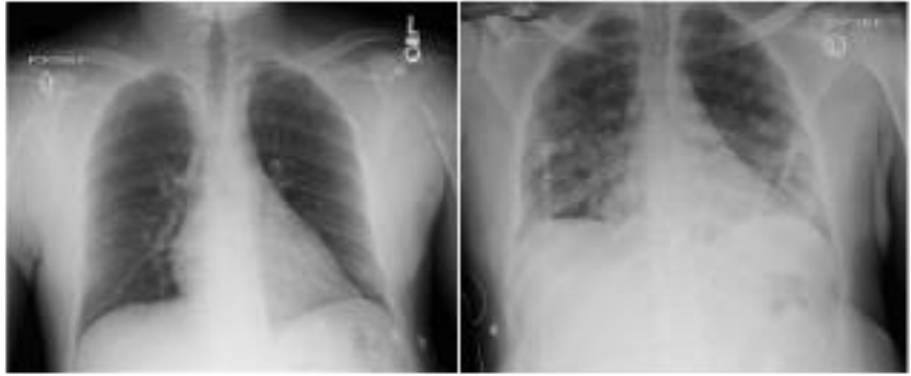

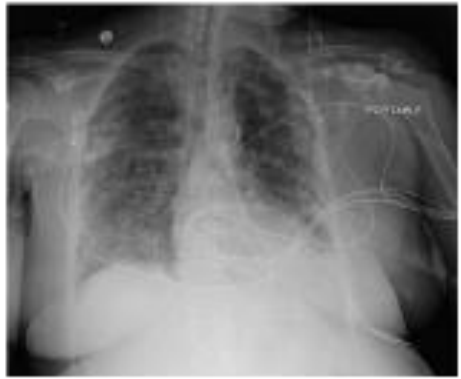

Figure 1: Overview of examples of different severity grades based on radiographic findings in chest X-rays: Normal (a): No clear sign of disease exists in the lung regions. Moderate (b): Bilateral opacification involving the peripheral mid and lower lung zones with approximately ≥ 25% and ≤ 50% involvement of the most severely affected lung. Severe (c): Patchy bilateral opacification with more than 50% involvement of the most severely affected lung.

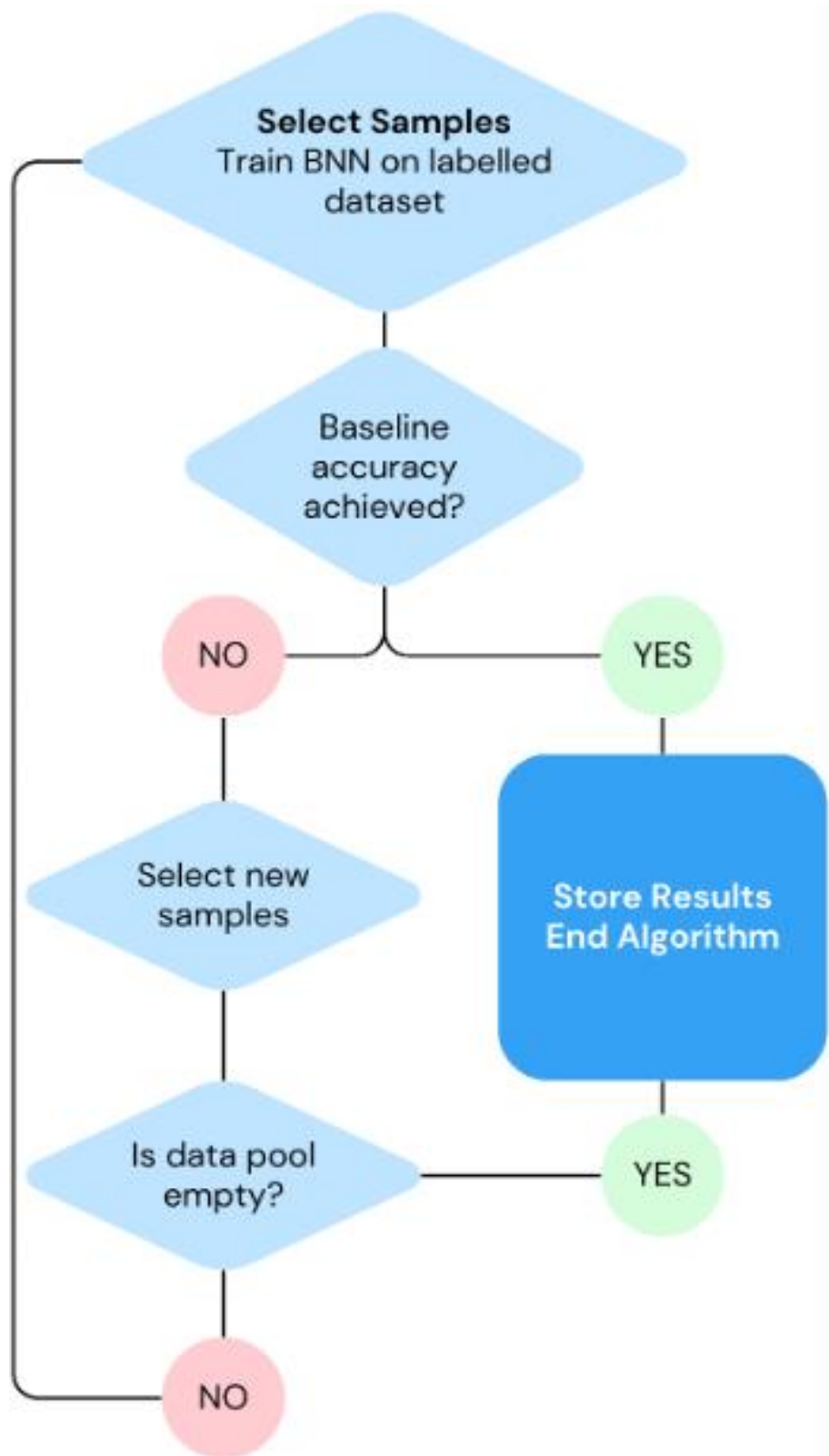

Figure 2: Overview of the active learning procedure. The process begins by selecting an initial batch of randomly chosen samples (25 per class severity) to train the Bayesian Neural Network (BNN). After training, we evaluate whether baseline accuracy has been achieved. If so, the result is stored, and the algorithm terminates. If not, a new batch of samples is selected based on the specified acquisition function (e.g., Entropy Sampling), and the BNN is retrained from scratch using the expanded labeled set. This iterative loop continues until either the data pool is exhausted, or baseline accuracy is reached.

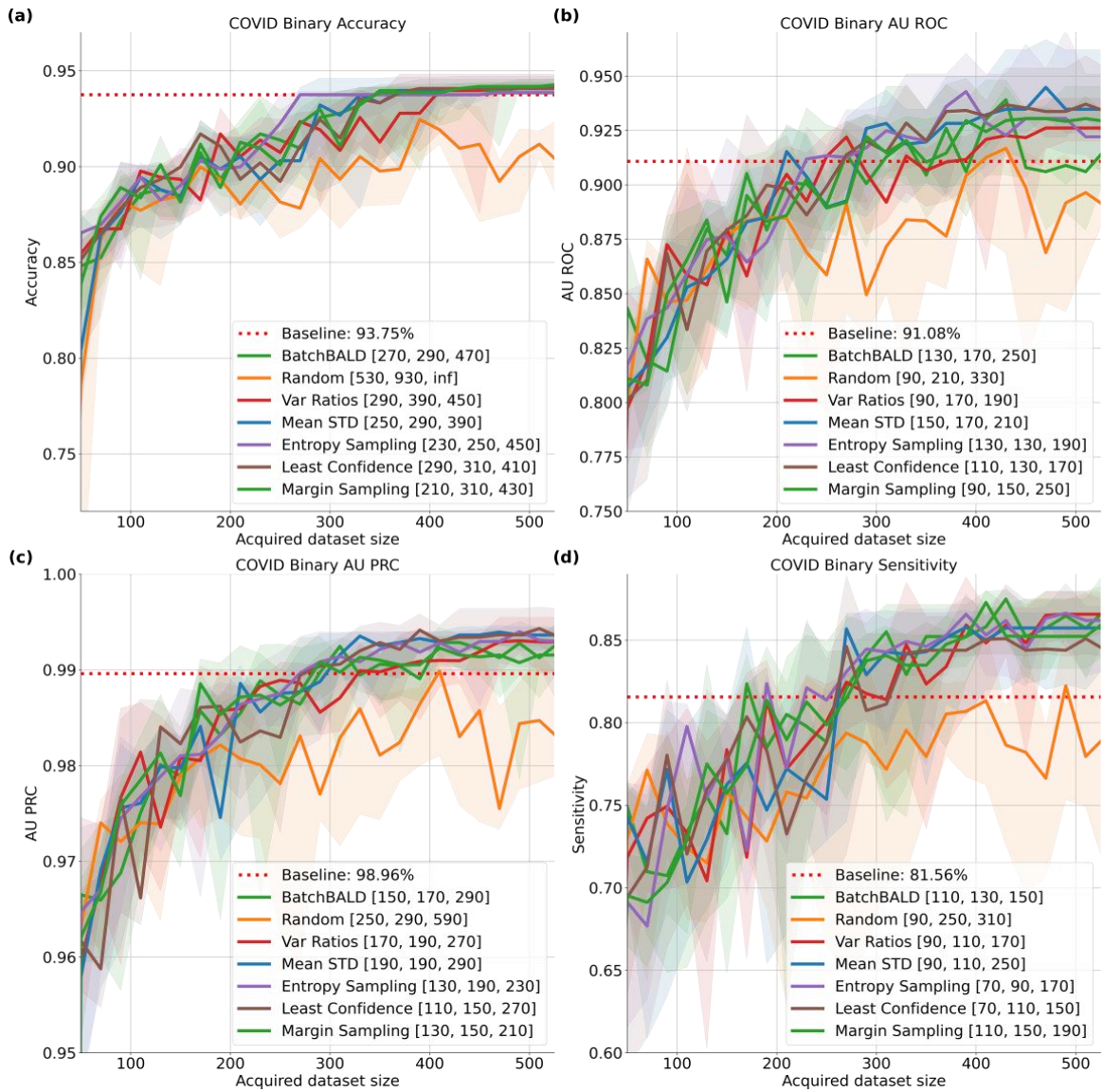

Figure 3: Binary classification results across various metrics as a function of the number of imaging data samples acquired: Accuracy (a), AU ROC (b), AU PRC (c), sensitivity (d). The results are reported as [25th, 50th, 75th] percentiles. The baseline is denoted as a dotted line in red.

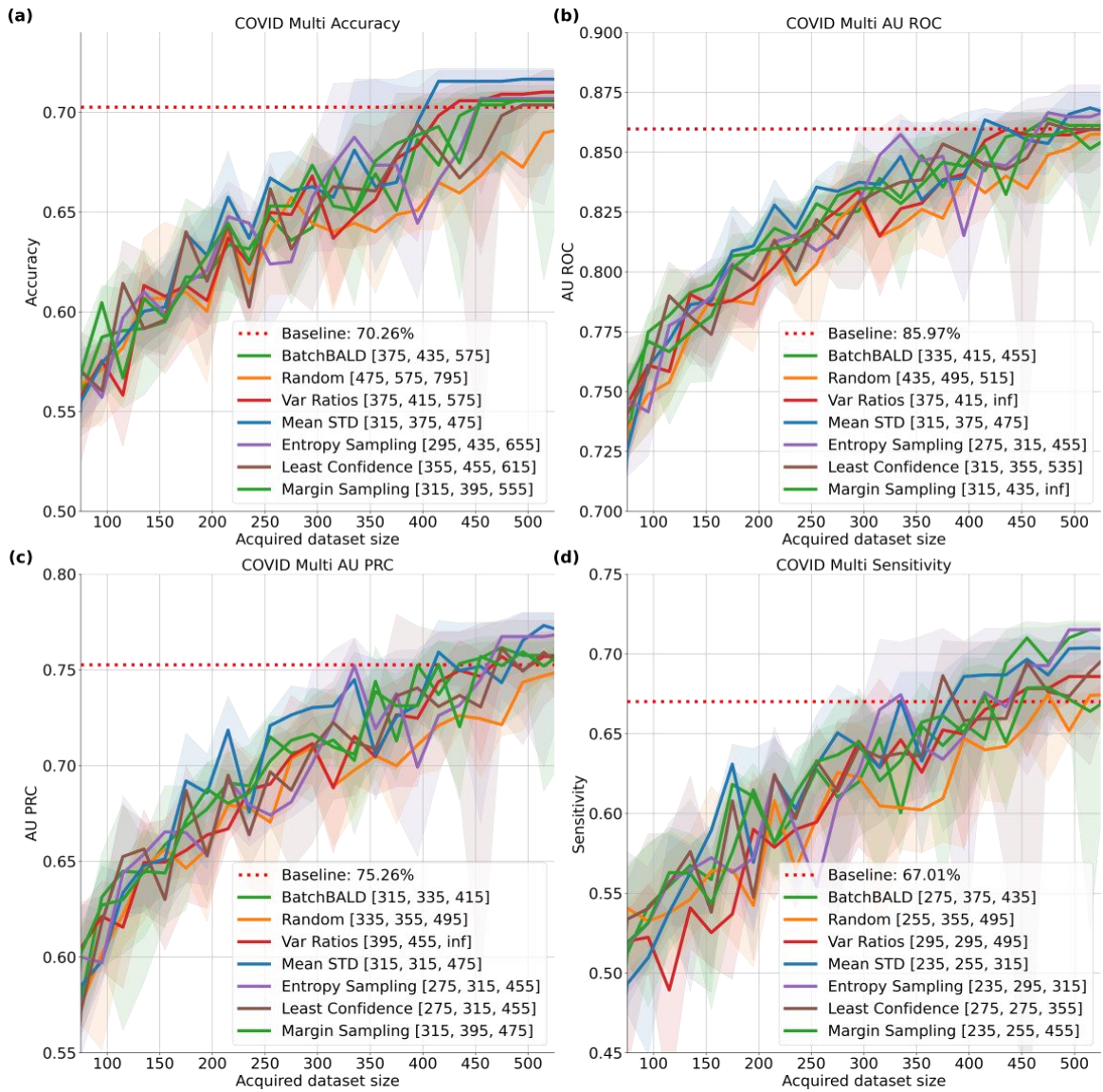

Figure 4: Multi-class classification results across various metrics as a function of the number of imaging data samples acquired: Accuracy (a), AU ROC (b), AU PRC (c), sensitivity (d). The results are reported as [25th, 50th, 75th] percentiles. The baseline is denoted as a dotted line in red.

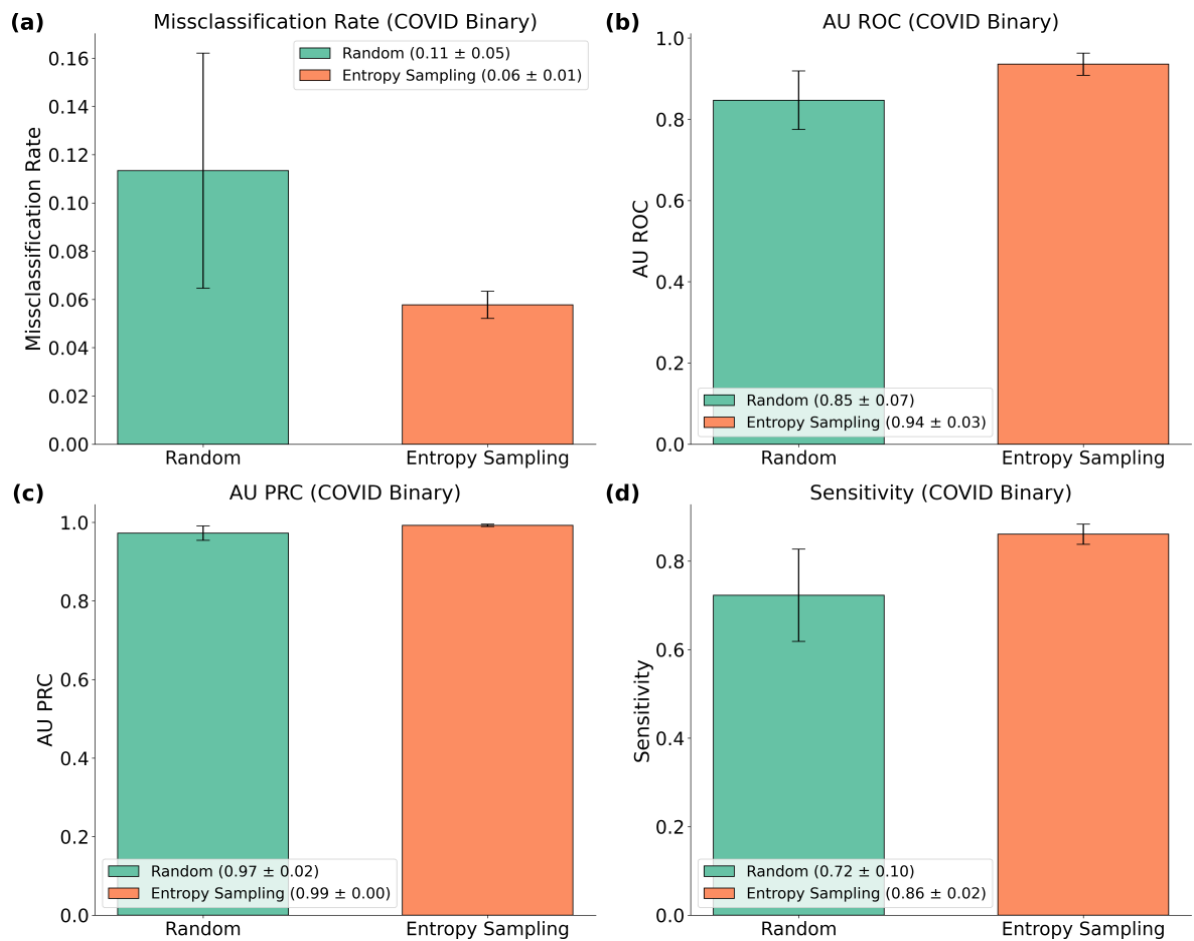

Figure 5: Binary classification error comparison between Entropy Sampling and Random Sampling using only 15.4% of the imaging data samples across various metrics: Misclassification Rate (a), AU ROC (b), AU PRC (c), sensitivity (d).

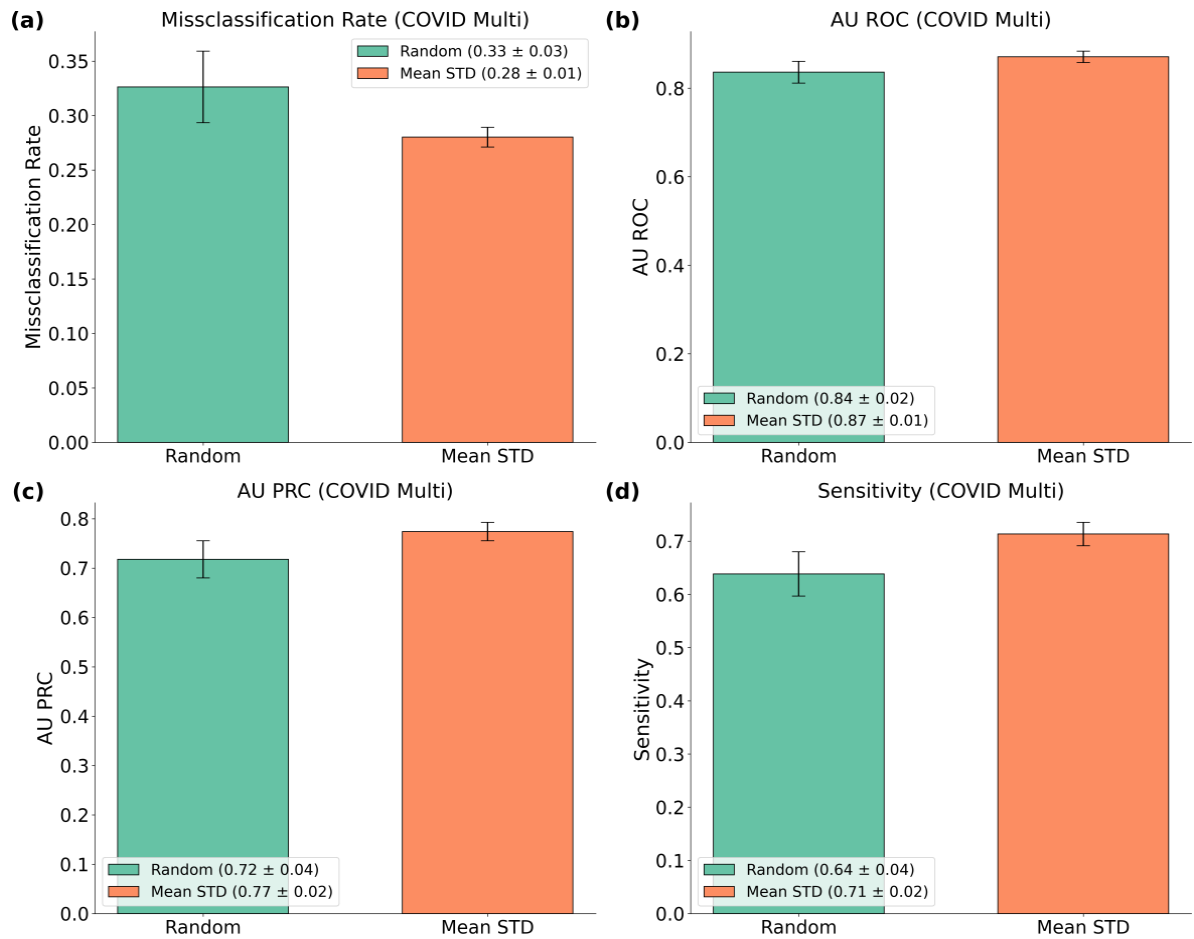

Figure 6: Multi-class classification error comparison between Mean STD and Random Sampling using only 23.1% of the imaging data samples across various metrics: Misclassification Rate (a), AU ROC (b), AU PRC (c), sensitivity (d).

**Supplementary Material Overview**

The supplementary material contains additional implementation details, acquisition function definitions, baseline tables, extended evaluation plots, and sampling/training time analysis not included in the main manuscript.

1. **Active Learning Implementation**

    The following definitions clarify the framework and notation used in this paper:

    - **Training Pool ($D_{pool}$):** This is the set of all unlabeled data points available for potential labeling. Formally, let $D_{pool} = \{x_1, x_2, \ldots, x_m\}$, where $x_i$ represents an individual data point in the pool.

    - **Training Set ($D_{train}$):** This is the set of labeled data points used to train the model. Initially, $D_{train}$ is populated with a predetermined number of labeled data points from each class. As active learning progresses, $D_{train}$ is updated with new labeled samples.

    - **Acquisition Function:** An acquisition function evaluates the informativeness of each data point in $D_{pool}$ to determine which points should be labeled next. For each acquisition function, the input is denoted by the vector $\mathbf{x}$, where $\mathbf{x} \in D_{pool}$.

    - **Selected Samples ($S_t$):** At each iteration $t$, an acquisition function selects a subset $S_t \subset D_{pool}$ of $n_t$ samples to be labeled. The size of $S_t$ and the value of $n_t$ is typically determined by the active learning strategy or a predefined budget.

    - **Updated Training Set ($D^t_{train}$):** After labeling the selected samples, the training set is updated to $D^t_{train} = D^{t-1}_{train} \cup S_t$, where $D^{t-1}_{train}$ is the training set before iteration $t$.

    The BNN was trained with the following specifics:

    - **Learning Rate:** A fixed learning rate of 0.001 is used throughout the training.
    - **Optimizer:** The Adam optimizer is utilized for its adaptive learning rate properties.
    - **MC Dropout:** Applied during training and inference to approximate Bayesian uncertainty.

- **Early Stopping:** Training stops when no improvement is seen over three successive iterations.
- **Initial Stratified Split:** Ensures the label distribution is consistent across all random seeds and initializations. This means that the distribution of class labels is uniform across different splits generated by the same random seed but may vary between different seeds. This approach maintains the inherent nature of the data distribution across different experiments.

2. **Acquisition Functions**

2.1. *Random Sampling*

Random Sampling selects data points uniformly at random from the unlabeled pool, $D_{pool}$. This method does not leverage any information from the model or data and serves as a baseline for comparing the effectiveness of more sophisticated acquisition functions.

2.2. *Entropy Sampling*

Entropy Sampling selects samples that maximize the entropy of the predictive distribution. Entropy, defined as:

$$H(\mathbf{x}) = -\sum_{c=1}^{C} \hat{p}(y = c \mid \mathbf{x}) \log \hat{p}(y = c \mid \mathbf{x})$$

is a measure of uncertainty, and this method chooses instances where the model's predictive distribution is most uncertain. The predictive probability, denoted as $\hat{p}(y = c \mid \mathbf{x})$, represents the BNN model's estimated probability that a data point **x** belongs to class $c$.

2.3. *Batch BALD*

BatchBALD extends the BALD method to batch mode active learning, where it jointly maximizes the mutual information between the model predictions and the model parameters across multiple points. This helps in selecting diverse and informative batches of data points.

The mutual information $I$ for a batch B is defined as:

$$I(\mathcal{B}; \theta \mid \mathcal{D}) = H(\mathcal{B} \mid \mathcal{D}) - \mathbb{E}_{\theta \sim p(\theta|\mathcal{D})}[H(\mathcal{B} \mid \theta, \mathcal{D})]$$

where $H$ is the entropy, $\theta$ are the model parameters, and D is the dataset.

### 2.4. Mean STD

Mean STD selects samples based on the standard deviation of the predictive probabilities. It aims to choose instances where the model is most uncertain, as indicated by a high standard deviation in the predicted class probabilities. This method is effective in highlighting regions of the input space where the model's predictions are less certain.

The standard deviation for a data point **x** is given by:

$$STD(\mathbf{x}) = \sqrt{\frac{1}{C} \sum_{c=1}^{C} (\hat{p}(y = c \mid \mathbf{x}) - \overline{\hat{p}(y = c \mid \mathbf{x})})^2}$$

where $C$ is the number of classes and $\hat{p}(y = c \mid \mathbf{x})$ is the mean predictive probability. In our approach, the mean predictive probability is calculated by averaging the predicted probabilities over multiple stochastic forward passes using MC Dropout to account for model uncertainty.

### 2.5. Least Confidence

Least Confidence selects samples where the model has the lowest confidence in its predictions. The confidence is typically defined as the highest class probability predicted by the model. This method targets instances where the model's most confident prediction is still relatively low.

The confidence score for a data point **x** is given by:

$$LC(\mathbf{x}) = 1 - \max_c \hat{p}(y = c \mid \mathbf{x})$$

where $\hat{p}(y = c \mid \mathbf{x})$ is the predicted probability for class $c$.

2.6. *Margin Sampling*

Margin Sampling selects instances based on the difference between the highest and second highest predicted probabilities. The margin is given by:

$$M(x) = \hat{p}(\hat{y} = c_1 \mid \mathbf{x}) - \hat{p}(\hat{y} = c_2 \mid \mathbf{x})$$

where $\hat{p}(\hat{y} = c_1 \mid \mathbf{x})$ and $\hat{p}(\hat{y} = c_2 \mid \mathbf{x})$ are the first and second highest predicted class probabilities, respectively. Smaller margins indicate higher uncertainty and thus are selected.

2.7. *Variation Ratios*

Variation Ratios is an uncertainty-based acquisition function that selects instances based on the variability in predictions across multiple model iterations. It is defined as:

$$VR(x) = 1 - \frac{1}{T} \sum_{t=1}^{T} \mathbb{I}[\hat{y}^t = \hat{y}^*]$$

where $\hat{y}^t$ is the predicted class label for the data point **x** during the $t^{th}$ forward pass, $\hat{y}^*$ is the mode of the predicted class label over the $T$ stochastic Monte Carlo forward passes, and $\mathbb{I}[\hat{y}^t = \hat{y}^*]$ is an indicator function that equals one if the prediction in the $t^{th}$ pass is equal to the most frequent prediction $\hat{y}^*$, and zero otherwise. This method captures the model's predictive uncertainty by considering the variability in class predictions.

3. **Tables and Figures**

Table 1: Baseline Performance Metrics on Binary Dataset

| Metric | Value |
|---|---|
| Accuracy | 0.9375 |
| NLL | 0.4527 |
| F1 | 0.8539 |
| Precision | 0.9099 |
| Recall/Sensitivity | 0.8156 |
| AU ROC | 0.9108 |
| AU PRC | 0.9896 |
| Specificity | 0.8630 |

Table 2: Baseline Performance Metrics on Multi Dataset

| Metric | Value |
|---|---|
| Accuracy | 0.7025 |
| NLL | 0.5565 |
| F1 | 0.6940 |
| Precision | 0.7351 |
| Recall/Sensitivity | 0.6701 |
| AU ROC | 0.8597 |
| AU PRC | 0.7526 |
| Specificity | 0.8236 |

Table 3: Class distribution (% of selected samples) reported as mean ± standard deviation for each acquisition function across binary and multi-class classification tasks. Corresponds to Figure 5.

| Acquisition Function | Binary: Normal / COVID | Multi: Normal / Moderate / Severe |
|---|---|---|
| BatchBALD | 32.2 ±21.6 / 67.8 ±21.6 | 30.8 ±18.1 / 41.4 ±14.1 / 27.9 ±15.9 |
| Entropy Sampling | 35.3 ±20.3 / 64.7 ±20.3 | 21.6 ±15.8 / 46.9 ±12.6 / 31.5 ±16.0 |
| Least Confidence | 35.7 ±20.6 / 64.3 ±20.6 | 22.1 ±14.9 / 46.9 ±13.0 / 31.0 ±15.0 |
| Margin Sampling | 36.5 ±21.0 / 63.5 ±21.0 | 17.7 ±10.2 / 47.0 ±11.5 / 35.3 ±12.5 |
| Mean STD | 34.4 ±22.1 / 65.6 ±22.1 | 30.5 ±20.9 / 41.8 ±15.7 / 27.8 ±18.2 |
| Random | 13.3 ±7.4 / 86.7 ±7.4 | 13.7 ±7.6 / 37.0 ±10.7 / 49.3 ±10.6 |
| Var Ratios | 35.0 ±20.2 / 65.0 ±20.2 | 22.7 ±15.3 / 47.2 ±13.5 / 30.1 ±14.5 |

Table 4: Training and acquisition (batch) time per iteration (in seconds), reported as mean ± standard deviation across binary and multi-class settings. Corresponds to Figure 6.

| Acquisition Function | Binary: Train / Batch Time (s) | Multi: Train / Batch Time (s) |
|---|---|---|
| BatchBALD | 270.0 ± 119.7 / 60.6 ± 9.6 | 266.5 ± 102.7 / 69.5 ± 11.1 |
| Entropy Sampling | 263.0 ± 112.8 / 44.3 ± 7.0 | 285.8 ± 129.6 / 41.3 ± 8.2 |
| Least Confidence | 263.1 ± 111.9 / 46.4 ± 6.8 | 308.7 ± 126.8 / 43.5 ± 5.6 |
| Margin Sampling | 279.4 ± 126.8 / 46.4 ± 5.8 | 279.2 ± 115.6 / 42.4 ± 7.0 |
| Mean STD | 245.4 ± 102.4 / 46.3 ± 5.8 | 271.9 ± 132.0 / 43.7 ± 6.4 |
| Random | 253.9 ± 109.9 / 0.0 ± 0.0 | 271.7 ± 120.5 / 0.0 ± 0.0 |
| Var Ratios | 277.8 ± 121.8 / 45.0 ± 6.3 | 287.8 ± 126.6 / 43.7 ± 6.1 |

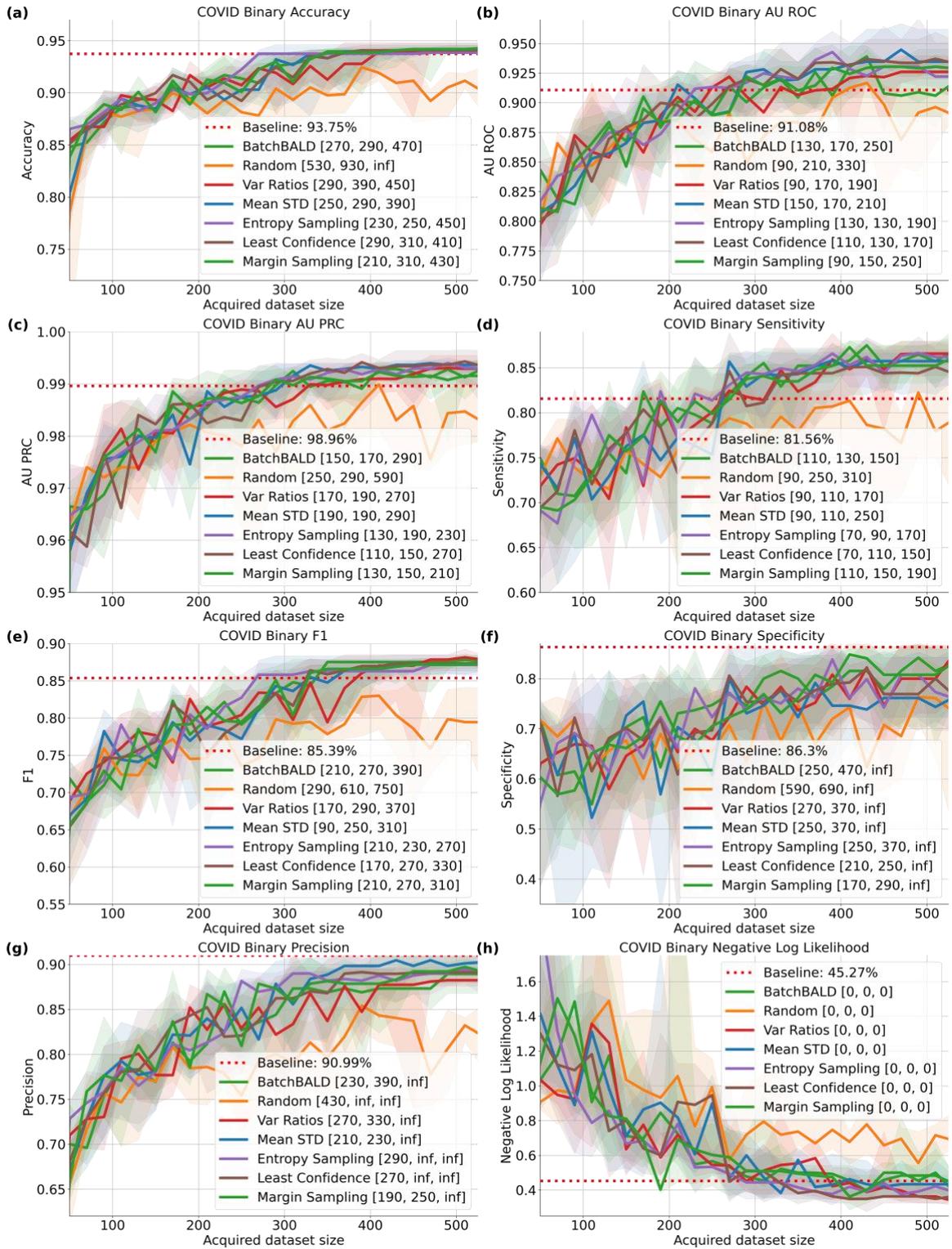

Figure 1: Binary classification results across various metrics as a function of the number of imaging data samples acquired: Accuracy (a), AU ROC (b), AU PRC (c), Sensitivity (d), F1 (e), Specificity (f), Precision (g), NLL (h). The results are reported as [25th, 50th, 75th] percentiles. The baseline is denoted as a dotted line in red.

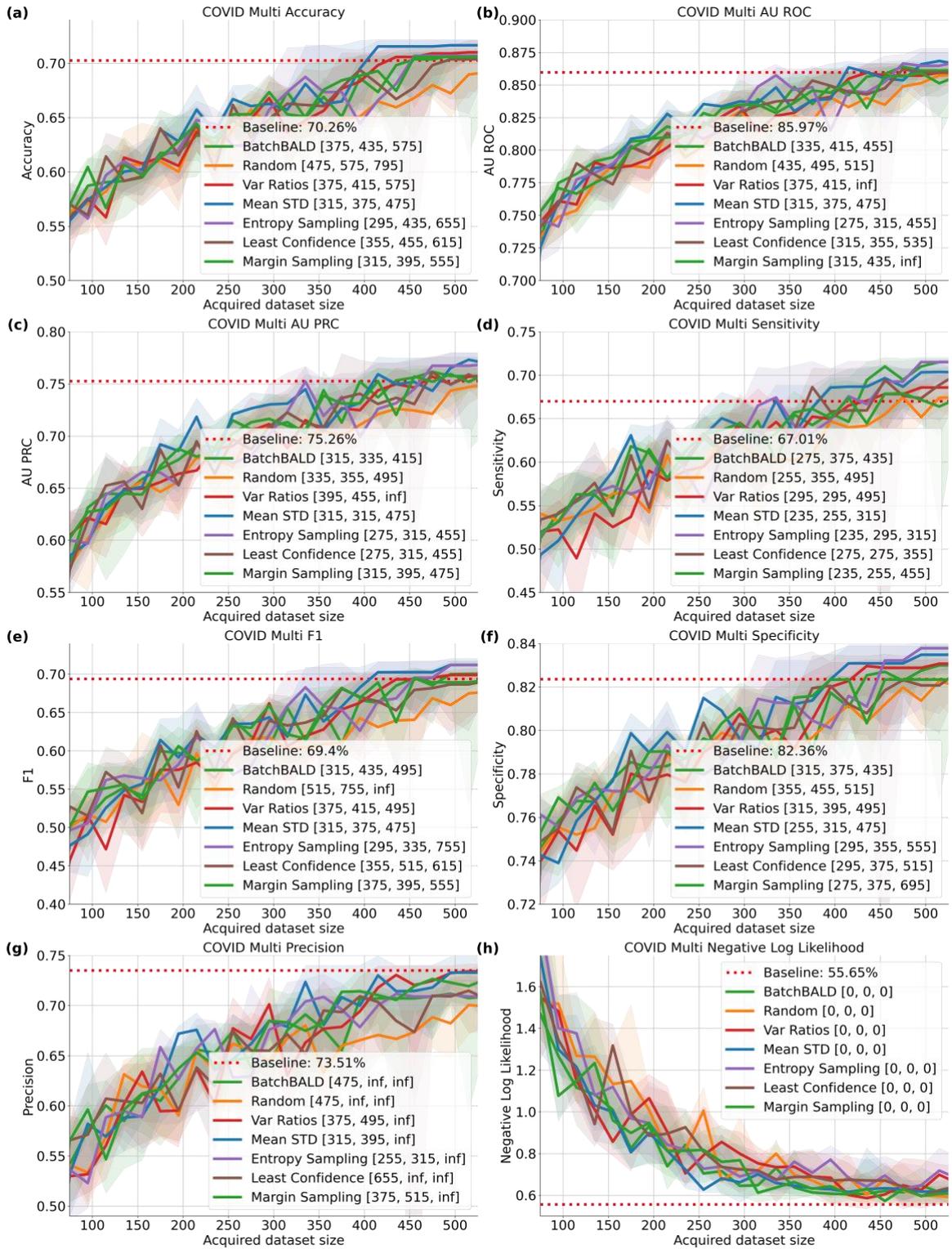

Figure 2: Multi-class classification results across various metrics as a function of the number of imaging data samples acquired: Accuracy (a), AU ROC (b), AU PRC (c), Sensitivity (d), F1 (e), Specificity (f), Precision (g), NLL (h). The results are reported as [25th, 50th, 75th] percentiles. The baseline is denoted as a dotted line in red.

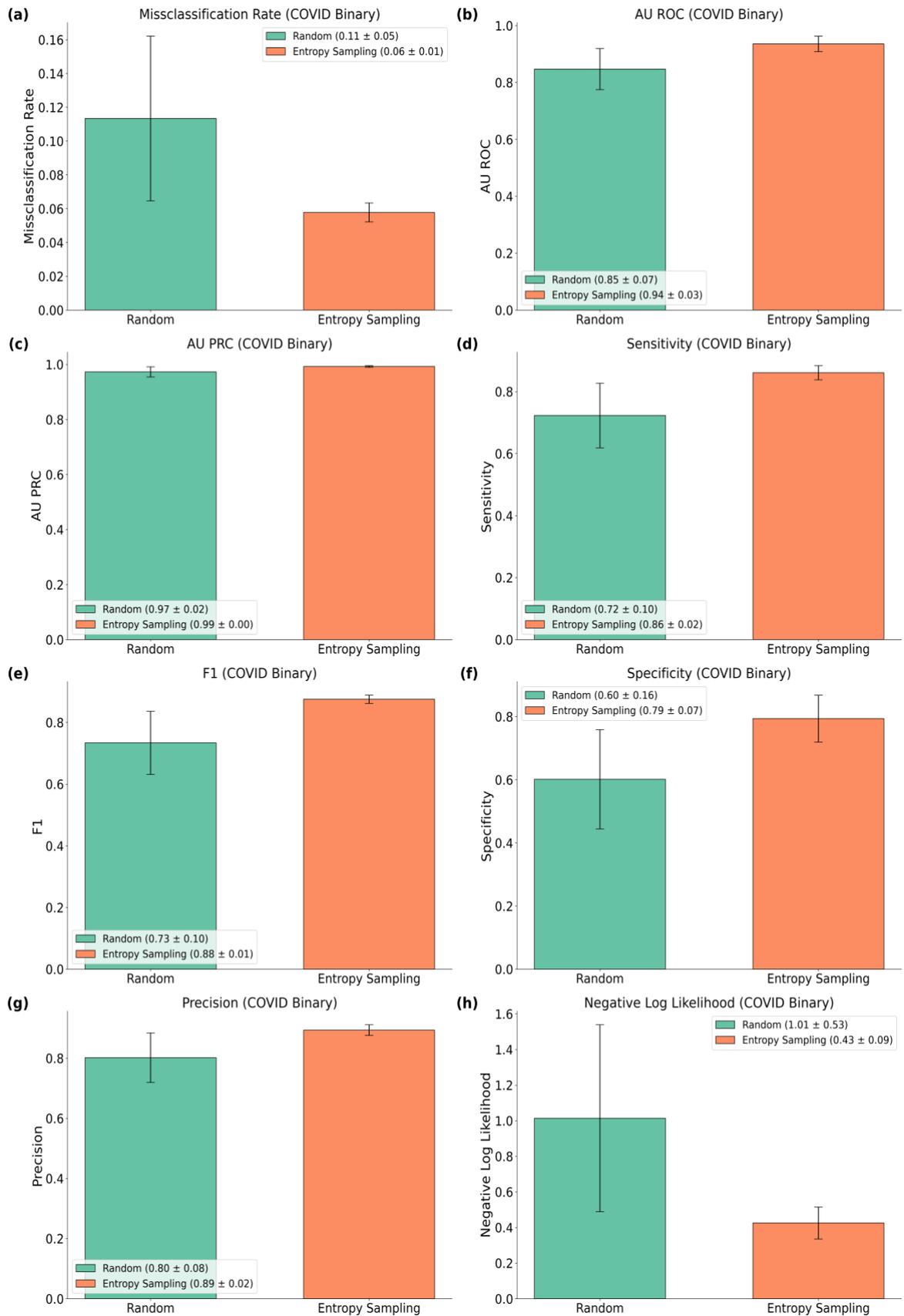

Figure 3: Binary classification error comparison between Entropy Sampling and Random Sampling using only 15.4% of the imaging data samples across various metrics: Misclassification Rate (a), AU ROC (b), AU PRC (c), Sensitivity (d), F1 (e), Specificity (f), Precision (g), NLL (h).

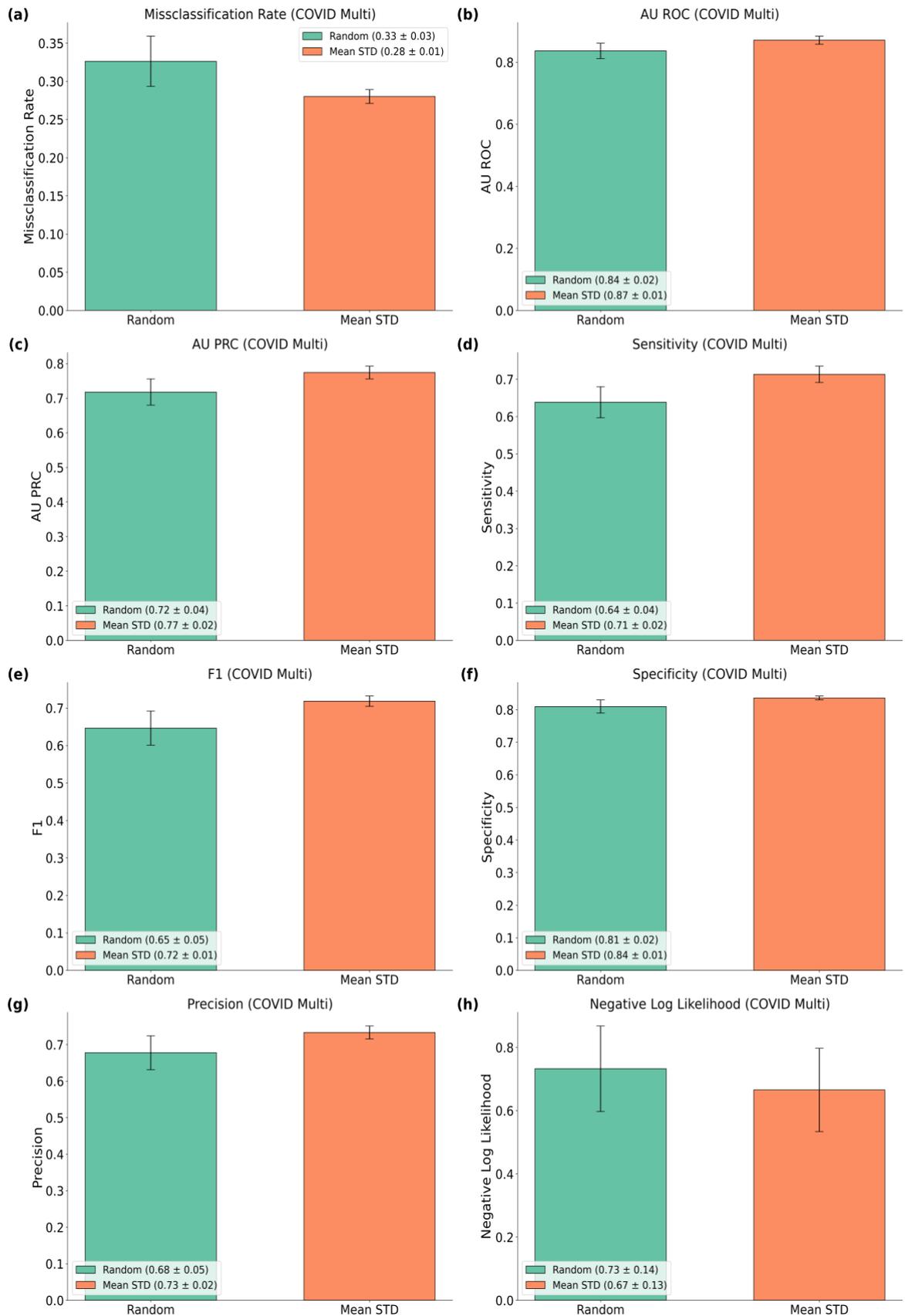

Figure 4: Multi-class classification error comparison between Mean STD and Random Sampling using only 23.1% of the imaging data samples across various metrics: Misclassification Rate (a), AU ROC (b), AU PRC (c), Sensitivity (d), F1 (e), Specificity (f), Precision (g), NLL (h).

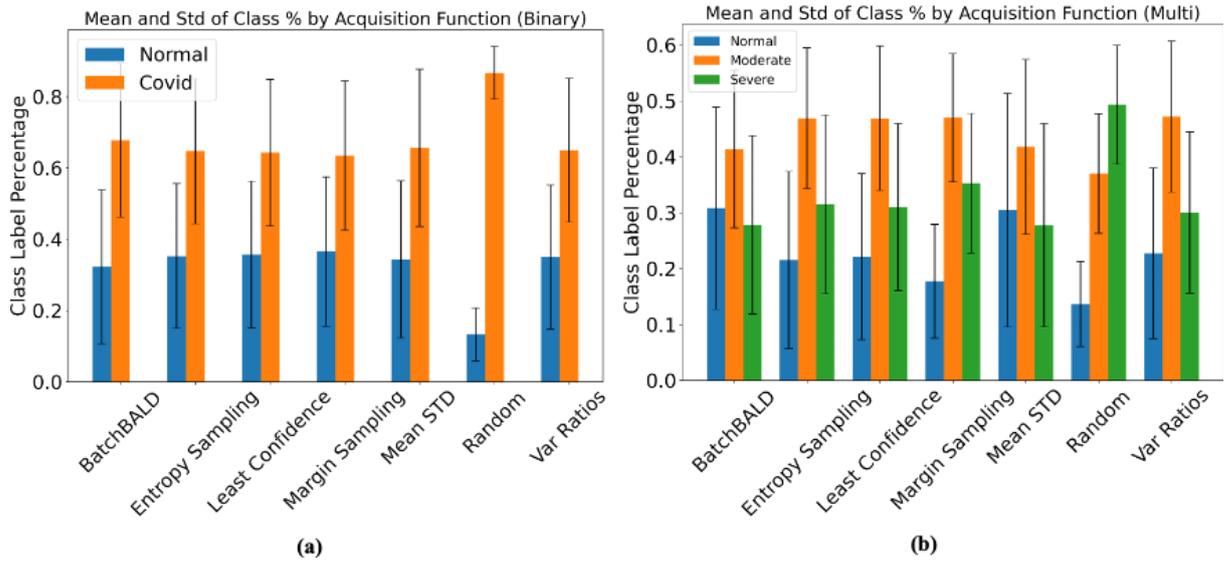

Figure 5: Mean and standard deviation of class distribution (in %) for binary (a) and multi-class (b) classification across various acquisition functions.

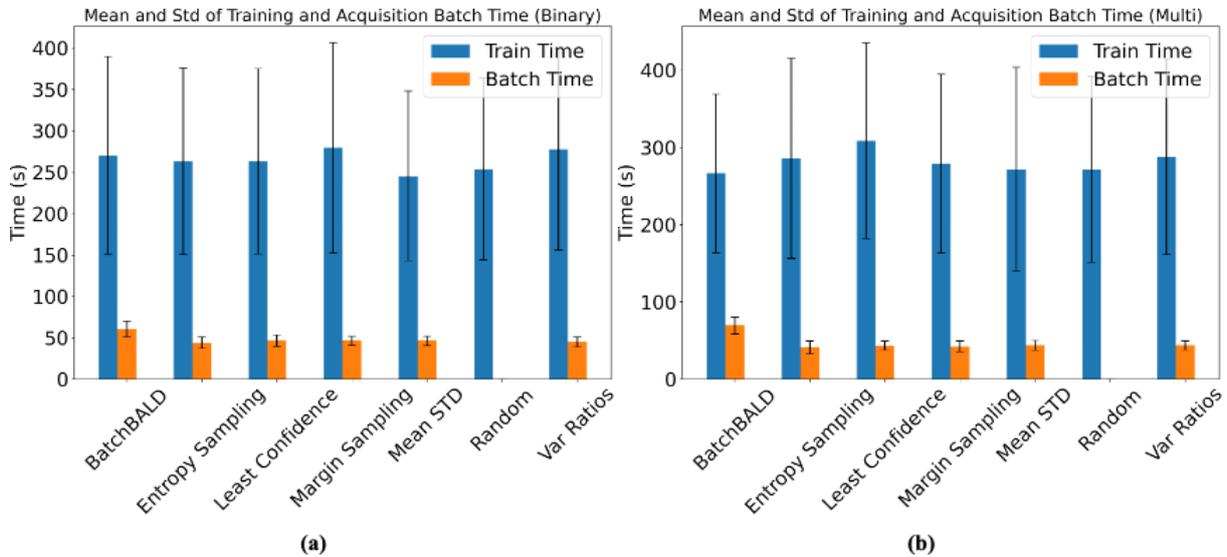

Figure 6: Mean and standard deviation of training and acquisition batch time for binary (a) and multi-class (b) classification across various acquisition functions.